\documentclass[twoside,12pt]{article}
\usepackage{amsmath,amsfonts,amssymb}
\usepackage{xy} \xyoption{all}
\usepackage{amsmath}
\usepackage{amsthm}
\usepackage{graphicx}
\usepackage[hang]{subfigure}
\usepackage{psfrag}
\usepackage{amscd}
\usepackage{url}
\newcommand{\goodgap}
{ \hspace{\subfigtopskip}
  \hspace{\subfigbottomskip} }

\makeatletter
\def\catchNameTitle{
\global\let\Title\@title }

\def\timenow{%
  \@tempcnta=\time \divide\@tempcnta by 60 \number\@tempcnta:\multiply
  \@tempcnta by 60 \@tempcntb=\time \advance\@tempcntb by -\@tempcnta
  \ifnum\@tempcntb <10 0\number\@tempcntb\else\number\@tempcntb\fi}

\def\TODAY{\ifnum\day<10 \number\day\else \number\day\fi\
\ifcase\month\or January\or February\or March\or April\or May\or
    June\or July\or August\or September\or October\or November\or December\fi,\ \number\year}

\newcommand{\MarkAll}[4]{
\renewcommand{\ps@myheadings}{
\renewcommand{\@oddhead}{{\scriptsize #1 \hfil #2}}
\renewcommand{\@evenhead}{\@oddhead}
\renewcommand{\@evenfoot}{{\scriptsize #3}\hfil
        \textrm{\thepage}\hfil{\scriptsize #4}}
\renewcommand{\@oddfoot}{\@evenfoot}}
}

\def\NOW{\TODAY \; at \timenow}

\newcommand{\DRAFT}{
\renewcommand{\@evenhead}{\hfil DRAFT \hfil}
\renewcommand{\@oddhead}{\@evenhead}
}

\newcommand{\NowFootNum}{
\renewcommand{\@evenfoot}{{\scriptsize ``\Title''\ by \ \Author }\hfil
         {\scriptsize \NOW}}
\renewcommand{\@oddfoot}{\@evenfoot}
}

\newcommand{\NowFoot}{
\renewcommand{\@evenfoot}{{\scriptsize \Title} {\scriptsize \NOW}}
\renewcommand{\@oddfoot}{\@evenfoot}
}

\newtheorem{Thm}{Theorem}[section]

\def\lb{_{\Lambda,\beta}}

\newtheorem{Prop}[Thm]{Proposition}
\newtheorem{Def}[Thm]{Definition}

\newtheorem{proposition}[Thm]{Proposition}

\newtheorem{lemma}[Thm]{Lemma}

\newtheorem{definition}[Thm]{Definition}
\def\eqbox#1{\fbox{$\displaystyle #1$}}

\def\Pref#1{Proposition~\ref{#1}}
\def\Cref#1{Corollary~\ref{#1}}
\def\Lref#1{Lemma~\ref{#1}}

\def\Dref#1{Definition~\ref{#1}}

\newtheorem{prop}{Proposition}

\def\gcd{{\rm g.c.d.}}
\def\calk{{\mathcal K}}

\def\calt{{\mathcal T}}

\def\caln{{\mathcal N}}

\def\cals{{\mathcal S}}

\def\calp{{\mathcal P}}

\def\calz{{\mathcal Z}}
\def\fz{{\frak Z}}

\def\fz{{\frak Z}}

\def\C{{\mathbb C}}
\def\Z{{\mathbb Z}}
\def\R{{\mathbb R}}

\def\pslash{\hbox{$\partial$\kern-1.2ex \raise.14ex\hbox{/}\kern.5ex}}
\def\Epslash{\hbox{{\rm {\hbox{$\partial$\kern-1.2ex \raise.14ex\hbox{/}\kern.5ex}\kern-.6ex \lower.28ex \hbox{$_E$}}}}}
\def\Epslashi{\hbox{{\rm {\pslash\kern-.6ex \lower.28ex \hbox{$_{E,i}$}}}}}
\def\Epslashii#1{\hbox{{\rm {\pslash\kern-.6ex \lower.28ex \hbox{$_{E,#1}$}}}}}
\def\Epslasht{\hbox{{\rm {\pslash\kern-.6ex \lower.28ex \hbox{$_{E,\calt}$}}}}}
\def\Epslashtstar{\hbox{{\rm {\pslash\kern-.6ex \lower.28ex \hbox{$_{E,\calt^*}$}}}}}
\def\Epslashti{\hbox{{\rm {\pslash\kern-.6ex \lower.28ex \hbox{$_{E,\calt_i}$}}}}}

\def\ep{\epsilon}

\def\hensp#1{\enspace\hbox{#1}\enspace}
\def\qsp#1{\qquad\hensp{#1}}

\def\l{\left}
\def\r{\right}

\def\part{\partial}

\def\be{\begin{equation}}
\def\ee{\end{equation}}
\def\beq{\begin{eqnarray}}
\def\eeq{\end{eqnarray}}
\def\nn{\nonumber \\ }

minus2pt \belowdisplayshortskip=12pt plus3pt minus2pt

\font\runningheadfont=cmcsc10

\def\lrp#1{\left( #1\right)}
\def\lrp#1{\left( #1\right)}
\def\lra#1{\left\langle #1\right\rangle}
\def\lraa#1{{\ll}{#1}{\gg}}

\def\abs#1{{\l\vert #1 \r\vert}}
\def\norm#1{{\l\Vert #1 \r\Vert}}

\newif\ifMarginNotes \MarginNotestrue
\def\mrgn#1{\ifMarginNotes\setbox0=\vtop{\hsize 6.75pc
   {\noindent\relax #1\par}}\leavevmode
   \vadjust{\dimen0=\dp0 \dimen1=\ht0\advance\dimen1 by .5ex
 \advance\dimen0 by -.5ex
  \kern-\dimen1\hbox{\kern\hsize\kern.5pc$\leftarrow$
  \box0}\kern-\dimen0}\fi}
\catcode`\@=11 \font\twelvemsb=msbm10 scaled 1200 \font\tenmsb=msbm10
\font\ninemsb=msbm7 scaled 1200
\newfam\msbfam
\textfont\msbfam=\twelvemsb  \scriptfont\msbfam=\tenmsb
  \scriptscriptfont\msbfam=\ninemsb
\def\msb@{\hexnumber@\msbfam}
\def\Bbb{\relax\ifmmode\let\next\Bbb@\else
 \def\next{\errmessage{Use \string\Bbb\space only in math
mode}}\fi\next}
\def\Bbb@#1{{\Bbb@@{#1}}}
\def\Bbb@@#1{\fam\msbfam#1}
   \font\twelveeufm=eufm10 scaled 1200 \font\teneufm=eufm10
 \font\seveneufm=eufm7 
\newfam\eufmfam
\textfont\eufmfam=\twelveeufm \scriptfont\eufmfam=\teneufm
\scriptscriptfont\eufmfam=\seveneufm
\def\frak{\relax\ifmmode\let\next\frak@\else
 \def\next{\errmessage{Use \string\frak\space only in math mode}}\fi\next}
\def\frak@#1{{\frak@@{#1}}}
\def\frak@@#1{\fam\eufmfam#1}

\catcode`\@=12

\title{Replica Condensation and Tree Decay}
\def\Author{Arthur Jaffe and David Moser}
\author{Arthur Jaffe\footnote{The authors thank an anonymous donor, whose
gift enabled this collaboration.} { and} David Moser\\
Harvard University\\
Cambridge, MA 02138, USA\\
{\small\urlstyle{same} \url{Arthur_Jaffe@harvard.edu},
\url{David.Moser@gmx.net}}}
\date{\today}

\begin{document}
  \catchNameTitle
  \maketitle
%  \hsize=6truein
%    \hoffset=-.75truein
%  \hsize=7truein \hoffset=-.75truein
%  \NowFootNum

\begin{abstract}
We give an intuitive method---using local, cyclic replica symmetry---to isolate
exponential tree decay in truncated (connected) correlations. We give an
expansion and use the symmetry to show that all terms vanish, except those
displaying {\em replica condensation}.  The condensation property ensures
exponential tree decay.

We illustrate our method in a low-temperature Ising system, but expect that one
can use a similar method in other random field and quantum field problems.
While considering the illustration, we prove an elementary upper bound on the
entropy of random lattice surfaces.
\end{abstract}

\tableofcontents

\section{Introduction}
Symmetry is used widely in physics to unify laws or simplify results. Global
symmetries often arise and are characterized by Lie groups or their
representation acting on a manifold.  Some symmetries, such as gauge symmetry,
are local; they are characterized by the action of a group on a bundle over a
manifold.  Global replica symmetry has been introduced as a symmetry of the
Hamiltonian of certain interacting systems such as Ising models, random fields,
and quantum fields, leading to valuable insights.

In \S \ref{Sect:Replicas} we study {\em local} replica symmetry.  This is {\em
not} a symmetry of the Hamiltonian in general, but it {\em is} a symmetry
within certain spin configurations.   This enables us to simplify our expansion
of certain expectations in the low-temperature Ising system in order to exhibit
a desired property: exponential tree decay of truncated correlations. This
low-temperature expansion only serves to illustrate our method. We plan to
investigate the use of our method in other high-temperature and low-temperature
situations for random and quantum fields.

Consider the truncated expectations
$\lra{\sigma_{i_1}\sigma_{i_2}\cdots\sigma_{i_n}}^{\rm T}$, defined in \S
\ref{Sect:TruncatedCorrelations}.  The Ising spins $\sigma_i$ are maps from the
unit lattice $\Z^d$ in $d\ge2$ dimensions to $\pm1$.  The Hamiltonian is
$H=\frac12\norm{\nabla\sigma}^2$,  and the Gibbs factor is $e^{-\beta H}$,
where $\beta$ denotes the inverse temperature. We show in \S
\ref{Sect:TreeDecay} that there are constants $a,b$ such that for $\delta_n=
\beta-b \ln n\ge 1$,
    \be
        \abs{\,\lra{\sigma_{i_1}\sigma_{i_2}\cdots\sigma_{i_n}}^{\rm T}}
        \le a\, n^n\, e^{- \delta_n\tau(i_1,\ldots,i_n)}\;,
    \label{FinalBound}
    \ee
where $ \tau(i_1,\ldots,i_n)$ is the length of the minimal tree connecting the
$n$ points $i_1,\ldots,i_n$.  Note the condition $\delta_n\ge1$ requires that
$\beta\ge\beta_n$, where $\beta_n$ grows at least as fast as $O(\ln n)$.  It
would be of interest to eliminate the $n$-dependence from the minimum value of
$\beta$.

Our method uses replica variables, comprising $n$ identical, independent copies
of the original system; one considers expectations in the replicated system
that are  product expectations for the individual systems. Replica symmetry is
the symmetry of these expectations under a permutation of the copies. For a
system in a finite volume $\Lambda$, with $i_1,\ldots,i_n\in\Lambda$, the same
estimate holds uniformly in $\Lambda$. Our method requires unbroken replica
symmetry, so one must impose the same boundary conditions in each replica copy.

We develop a low-temperature expansion, based on the intuitive idea that
individual terms with less than the desired exponential tree-graph decay sum to
zero (vanish) due to symmetry under the local cyclic replica group. In \S
\ref{Sect:TreeDecay} we define and establish convergence of this expansion. The
terms in the expansion are parameterized by {\em replica continents}.   These
replica continents are bounded by {\em random surfaces}. The convergence of our
expansion relies on an interplay between energy and entropy estimates; in
particular we give entropy estimates bounding the number of random surfaces
that occur in our expansion, as well as energy estimates showing that large
islands are suppressed at a desired rate.

Key to our method is the use of local cyclic replica symmetry, to show that all
non-zero terms in our expansion display {\em replica condensation}, defined in
\S\ref{Sect:ReplicaCondensation}.  By this we mean that all the lattice sites
$i_1,\ldots,i_n$ must live on a single continent. The size of the boundary of
the continent must therefore be larger than $\tau(i_1,\ldots,i_n)$; this is the
source of the exponential tree decay.

\subsection{The Ising Model as Illustration} The Ising system is the simplest example of a
statistical mechanics interaction.  We present our method for such a model on a
unit cubic lattice $\Z^d$, with $d\ge2$, although our methods clearly  apply in
more generality. The Ising Hamiltonian in volume $\Lambda\subset\Z^d$ is
    \be
        H_\Lambda
        = H_\Lambda(\sigma)
        = \frac12 \norm{\nabla \sigma}^2_{\ell^2(\Lambda)}
        = \sum_{nn\in\Lambda} \lrp{1-\sigma_i\sigma_j}\;,
    \label{IsingHamiltonian}
    \ee
where $nn$ denotes the sum over nearest-neighbor pairs of sites in the lattice,
namely sites with $\abs{i-j}=1$.  The partition function
    \be
        \calz_{\Lambda,\beta}
        = \sum_{{\sigma_i}\atop i\in\Lambda} e^{-\beta H_\Lambda(\sigma)}
    \label{NormalPartitionFunction}
    \ee
normalizes statistical averages $\lra{f}_{\Lambda,\beta}$ of a function $f$,
namely
    \be
        \lra{f}_{\Lambda,\beta}
        = \frac{1}{\calz_{\Lambda,\beta}}\,{\sum_{{\sigma_i}\atop i\in\Lambda}
             f(\sigma)\, e^{-{\beta H_\Lambda(\sigma)} }}\;.
    \label{NormalExpectation}
    \ee
Often $f$ is a monomial in spins,
$f=\sigma_{i_1}\sigma_{i_2}\cdots\sigma_{i_n}$.  The expectation $\lra{\ \cdot\
}\lb$ is linear, so one can express the expectation of a general $f$ as a limit
of finite linear combinations of expectations of the form
$\lra{\sigma_{i_1}\sigma_{i_2}\cdots\sigma_{i_n}}_{\Lambda,\beta}$.

\setcounter{equation}{0}
\section{The Correspondence $\Z^d\leftrightarrow\R^d$
\label{Sect:LatticeConnectedness}} 
\begin{figure}[h]
   \[
   \begin{array}{ccc}
       \Z^2 & & \R^2 \vspace {2mm} \\
       \includegraphics[height=2.5cm]{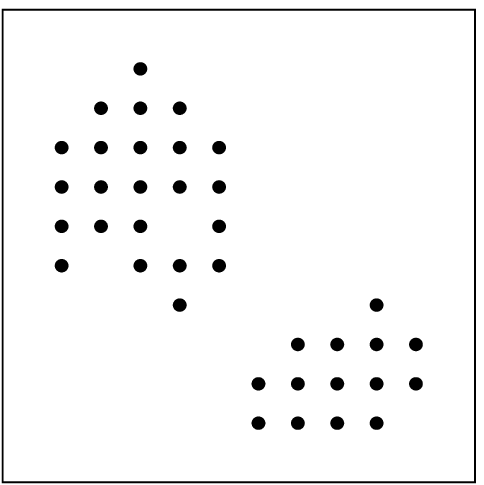}
          & \raisebox{1.15cm}{$\longleftrightarrow$}
              & \includegraphics[height=2.5cm]{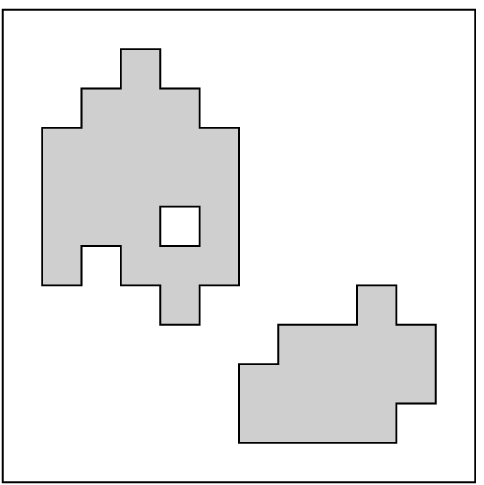} \vspace {2mm} \\
       \begin{CD} @V{\partial}VV \end{CD} && \begin{CD} @V{\partial}VV \end{CD} \vspace {2mm} \\
       \includegraphics[height=2.5cm]{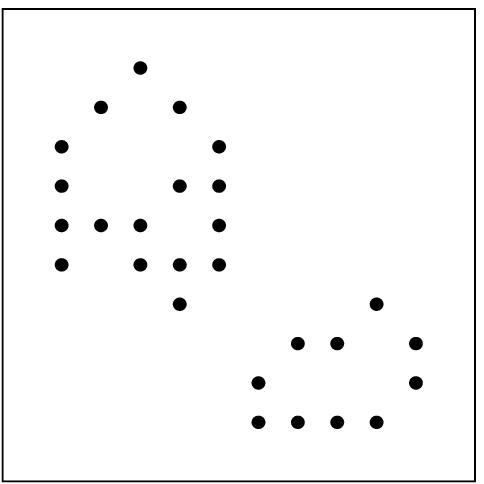}
          &
              & \includegraphics[height=2.5cm]{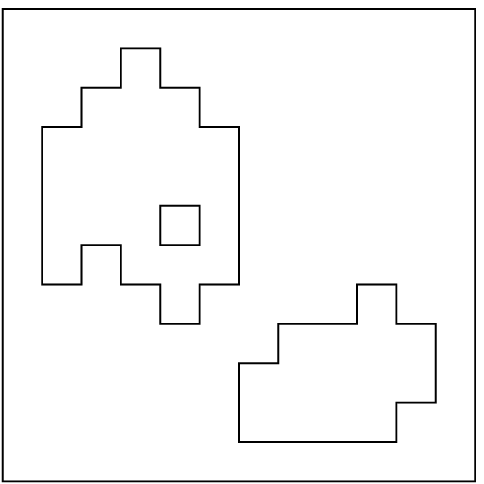} \\
   \end{array}
   \]
\caption{An example for the correspondence between subsets of $\Z^d$ and
$\R^d$, and their boundaries. \label{Figure:Correspondence} }
\end{figure}

Each subset $X\subset\Z^d$ of sites in the
lattice $\Z^d$ can be identified with a subset $X\subset\R^d$. Define the
latter as the union of closed, unit $d$-cubes $\square_i$ centered at the
lattice sites $i\in X$, as we illustrate in the upper part of Figure
\ref{Figure:Correspondence}.

\paragraph{Connectedness:}  We say that $X\subset\Z^d$ is {\em connected} if any two sites in $X$ can be
connected by a continuous path through nearest-neighbor lattice sites in the
set $X$.  This agrees with the notion that the interior of the  set
$X\subset\R^d$ is  connected in the ordinary sense. Two cubes are connected if
they share a $(d-1)$-dimensional face, but they are disconnected if they only
touch on a corner of dimension $\le (d-2)$.

\paragraph{Boundary:}
The boundary $\partial X\subset\R^{d}$ allows us to define the set $\partial
X\subset \Z^d$ of boundary lattice sites.  These boundary sites $\partial
X\in\Z^d$ are those lattice sites in $X$ lying in cubes that share a
$(d-1)$-dimensional face with the boundary $\partial X\subset \R^d$.

By $\abs{\partial X}$ we always refer to the area of the $(d-1)$-dimensional
surface in $\R^d$ and not the number of points in $\Z^d$. (A single cube
$\square_i$, for example, contains exactly $1$ boundary lattice site, while
$\abs{\,\square_i}=2d$.) In most instances we will call this area the
``length'' of the boundary, but in some cases we will also call it the number
of faces of the boundary surface. We illustrate the correspondence between the
boundary lattice sites and the boundary of regions in $\R^d$ in the lower part
of Figure~\ref{Figure:Correspondence}.

\paragraph{Surface:} More generally let a {\em face} in $\R^d$  denote a
$(d-1)$-cube; such a cube lies in the boundary of two $d$-cubes in $\R^d$. A
{\em surface} $Y$ is a union of $(d-1)$-faces, and its area $\abs{\;Y}$ is the
number of $(d-1)$-faces in $Y$.  Lattice sites in $Y$ may lie on either side of
the surface $Y$, but could be limited by selecting an orientation to
appropriate sets of faces in $Y$.

\paragraph{Connected Surface:} Define two faces to be adjacent, if they share a $(d-2)$-cube.
Likewise, define $Y$ to be {\em connected} if any two faces in $Y$ can be
reached by a continuous path through a sequence of adjacent faces in $Y$.

\setcounter{equation}{0}
\section{Replica Variables and Symmetry\label{Sect:Replicas}}
Choose $n\in\Z_+$ and consider $n$ independent copies of a
statistical-mechanical or quantum-field system; these are called $n$ replicas.
One can study the properties
of expectations under the group of permutations of the replica variables (the
{\em replica group}).  The $n$-element subgroup of cyclic permutation of all
the copies is abelian, and it provides useful one-dimensional representations
of replica symmetry.

\subsection{Replica Variables}
We assume that the different replicas are identical and independent.  They are
defined on the same lattice, they have the same form of interaction, they are
given identical boundary conditions, etc.  We label the spin variable at the
lattice site $i$ by $\sigma^{(\alpha)}_i$, where $\alpha = 1,2,\ldots,n$
denotes the index of the copy.  We also consider the replica spins at site $i$
as a vector $\vec \sigma_i$ with the vector components $\sigma_i^{(\alpha)}$.

\subsection{The Global Replica Group \label{Section:GlobalReplicaGroup}}
The {\em global replica group} is the symmetric group $S_n$
comprising elements $\pi\in S_n$ with action,
    \be
       \pi: (1,\ldots,n) \mapsto ({\pi_1},\ldots,{\pi_n})\;.
    \ee
The element $\pi\in S_n$ acts on the spins, giving a unitary
representation,
    \be
        \eqbox{\sigma_{i}^{(\alpha)}
        \mapsto
        \lrp{\pi\sigma_{i}}^{(\alpha)}
        = \sigma_{i}^{({\pi^{-1}}_{\alpha})} }\;,
        \qsp{for}
        \alpha=1,\ldots,n\;,\hensp{and for all }i \;.
    \label{GlobalReplicaGroup}
    \ee
The {\em global cyclic replica group} $S^c_n$ is the subgroup of cyclic
permutations of $n$ objects, and is generated by the permutation $\pi^0$,
    \be
       \pi^0: (1,\ldots,n) \mapsto ({2},\ldots,{n},1)\;.
    \ee
Treating the indices $\alpha$ modulo $n$, substitute $\alpha=n$ for $\alpha=0$
and write
    \be
        \eqbox{\sigma^{(\alpha)}_i
        \mapsto \lrp{\pi^0 \sigma_i}^{(\alpha)}= \sigma^{(\alpha-1)}_i}\;,
        \qsp{for}
        \alpha = 1,\ldots,n\;,\hensp{and for all }i \;.
    \label{ReplicaSpinTransformation}
    \ee
The matrix representation of  \eqref{ReplicaSpinTransformation} is
$\vec\sigma_i\mapsto\pi^0\vec\sigma_i$, where
    \be
        \lrp{\pi^0\vec\sigma_i}^{(\alpha)}
        = \sum_{\alpha'=1}^n
        \lrp{\pi^0}_{\alpha\,\alpha'}\sigma_i^{(\alpha')}\;,
        \qsp{and}
        \lrp{\pi^0}_{\alpha\alpha'}
        = \delta_{\alpha-1\ \alpha'}
        \;.
    \label{CyclicGenerator}
    \ee

\subsection{The Local Cyclic Replica Group \label{Section:LocalReplicaGroup}}
Let $\calk$ denote a subset of the lattice $\Z^d$.  The {\em local cyclic
replica group} $ S^{c}_n(\calk)$ is a bundle over $S_n^c$ defined as the action
of $S_n^c$ on the spins in $\calk$ and the identity on the complement.  This
group is generated by $\pi^0_\calk$ which has the representation on spins,
    \be
        \pi^0_\calk\vec\sigma_{i}
        = \l\{\begin{matrix}
            \pi^0\vec\sigma_{i}\;,& \hensp{when }i\in\calk\hfill\\
            \vec\sigma_{i}\;, &\hensp{when }i\not\in\calk\hfill
            \end{matrix} \r.      \;.
    \label{LocalCyclicReplicaTransformation}
    \ee

\subsection{Irreducible Representations}
The cyclic replica group is abelian, so its irreducible representations are one
dimensional.  We transform from $\vec\sigma_i$ to a set of coordinates $\vec
s_i=U\vec\sigma_i$ to reduce the representation of $S_n^c$.  In particular, let
$\omega=e^{2\pi i/n}$ denote the primitive $n^{\rm th}$ root of unity. Define
    \be
        \eqbox{s^{(\alpha)}_i
        = \frac{1}{n^{1/2}}\sum_{\alpha'=1}^{n} \omega^{\alpha(\alpha'-1)}
        \sigma^{(\alpha')}_i}\;,
        \qsp{for}
        \alpha = 1,\ldots,n\;.
    \label{ReplicaSums}
    \ee
Note that for $n>2$ the $s$-variables may be complex, even though the original
$\sigma$-spins are real.  The choice \eqref{ReplicaSums} defines the entries of
the matrix $U$ as  $U_{\alpha\alpha'}=n^{-1/2}\omega^{\alpha\lrp{\alpha'-1}}$.
This is Fourier transform in the replica space.
\begin{proposition} \label{Prop:CyclicReplicaSymmetryDiagonal}
The matrix $U$ is unitary with eigenvalues $\omega^\alpha$, for
$\alpha=1,\ldots,n$. Let $D$ be the diagonal matrix with $D_{\alpha\alpha'} =
\omega^{\alpha}\delta_{\alpha\alpha'}$.  Then
    \be
         \pi^0\vec s_i = D\vec s_i\;.
    \label{DiagonalPi}
    \ee
\end{proposition}

\begin{proof}[\bf Proof]
For $\nu$ an integer (modulo $n$),
    \be
        \sum_{\alpha=1}^n \omega^{-\nu\alpha}=n\,\delta_{\nu0}\;.
    \label{RootsOfUnitySum}
    \ee
Thus
    \be
        \lrp{UU^*}_{\alpha\alpha'}
        = \sum_{\beta=1}^n U_{\alpha\beta} \overline{U_{\alpha'\beta}}
        = \frac{1}{n}\sum_{\beta=1}^n \omega^{(\alpha-\alpha')(\beta-1)}
        = \delta_{\alpha\alpha'}\;.
    \ee
Since $\pi^0$ acts on the $\vec\sigma_i$ components according to
\eqref{ReplicaSpinTransformation}, this means that
    \be
        \lrp{\pi^0\vec s_i}^{(\alpha)}
        = \omega^{\alpha} \lrp{\vec s_i}^{(\alpha)}
        = \sum_{\alpha'=1}^n  D_{\alpha\alpha'}\lrp{\vec s_i}^{(\alpha)}\;,
    \ee
which is \eqref{DiagonalPi}.
\end{proof}

The inverse change of coordinates is
    \be
        \eqbox{\sigma^{(\gamma)}_i
        = \frac{1}{n^{1/2}}\sum_{\alpha=1}^n  \omega^{-(\gamma-1)\alpha}
        s^{(\alpha)}_i}\;,
        \qsp{for}
        \gamma = 1,\ldots,n\;.
     \ee
A further corollary of the unitarity of $U$ is the fact that for any $i,j$
    \be
        \sum_{\alpha=1}^n {\sigma^{(\alpha)}_i \sigma^{(\alpha)}_j}
        = \lra{\vec \sigma_i , \vec \sigma_j}_{\ell^2}
        = \lra{{{U\vec\sigma_i}},  {U\vec\sigma_j}}_{\ell^2}
        = \lra{\vec s_i, \vec s_j}_{\ell^2}
        = \sum_{\alpha=1}^n\overline{s_{i}^{(\alpha)}}s_j^{(\alpha)}\;.
    \label{SpinTimesSpin}
    \ee
In particular, the expression on the right side of this identity is always
real.  Furthermore, each individual term on the right is invariant under the
elements of the local, cyclic replica group $ S^{c}_n(\calk)$ as long as both
$i,j\in\calk$ or both $i,j\not\in\calk$.

\subsection{Replica Boundary Conditions}
We consider finite volume Hamiltonians that, along with their boundary
conditions, have the global replica group as a symmetry.    If one wished to
investigate the breaking of the replica group in the infinite volume limit,
then one might explicitly break replica symmetry in a finite volume by imposing
different boundary conditions for different replica copies of the system.

Since our system is originally given in terms of the variables $\sigma_i$, one
describes the boundary conditions in the volume $\Lambda$ in terms of the
variables $\sigma_i$ for $i\in\partial\Lambda$, with $\partial \Lambda$
defined in \S \ref{Sect:LatticeConnectedness}.

It is natural to ensure symmetry under the replica group by specifying the same
boundary condition on each component of the vector spin
    \be
        \sigma^{(\alpha)}_i
        = \sigma_i\;,
        \qsp{for all}
        i\in\partial\Lambda\;,
        \hensp{and all}
        \alpha=1,\ldots,n\;.
    \ee
In order to simplify the discussion, we impose $+1$ boundary conditions in each
replica copy:  set
    \be
        \eqbox{\vec\sigma_i
        = (+1,\ldots,+1)}
        \;,
        \hensp{when}i\in\partial \Lambda\;.
    \ee
The resulting boundary conditons for   $\vec s$ are
    \be
        \vec s_{i}
       = \lrp{0,0,\ldots,0,n^{1/2}}        \;,
        \hensp{when}i\in\partial \Lambda\;.
    \label{SBoundaryConditions}
    \ee

\subsection{Replica Symmetry is Global, not Local
\label{Section:BasicReplicaSymmetry}}
Define the total replica Hamiltonian $H_{\rm replica}$ as the sum of the
Hamiltonians for the replica copies of the Hamiltonian in volume $\Lambda$,
    \be
        \eqbox{H_{\rm replica}
        = H_{\rm replica}(\vec\sigma)
%        = \sum_{\alpha=1}^n H_\Lambda(\sigma^{(\alpha)})
        = \frac12 \norm{\nabla \vec \sigma}_{\ell^2(\Lambda)}
        =  \frac12\sum_{\alpha=1}^n\,\sum_{{\rm nn \in\Lambda }}
            \lrp{\sigma_{i}^{(\alpha)} - \sigma_j^{(\alpha)}}^2}\;.
    \label{MultipleReplicaHamiltonianSigma}
    \ee

\begin{proposition} \label{Prop:GlobalReplicaSymmetryH} Consider the replica Hamiltonian
\eqref{MultipleReplicaHamiltonianSigma}.
    \begin{itemize}
    \item[i.] As a function of the variables $\vec s$, one has
    \be
        \eqbox{H_{\rm replica}
        = \frac12 \norm{\nabla \vec \sigma}_{\ell^2(\Lambda)}
        = \frac12 \norm{\nabla \vec s}_{\ell^2(\Lambda)}
        = \frac12\sum_{\alpha=1}^n \;\sum_{{\rm nn \in\Lambda }}
        \abs{s_{i}^{(\alpha)} - s_j^{(\alpha)}}^2}\;.
    \label{MultipleReplicaHamiltonian}
    \ee
    \item[ii.] The replica Hamiltonian
    \eqref{MultipleReplicaHamiltonian} is invariant under a global replica permutation
    $\pi\in S_n$ defined in \eqref{GlobalReplicaGroup}, namely
    \be
       \eqbox{ H_{\rm replica}(\pi\vec s) = H_{\rm replica}(\vec s)}\;.
    \label{ReplicaInvariance}
    \ee
    \item[iii. ]  In general, the replica Hamiltonian is {\em not} invariant under the local
cyclic replica group $S^c_n(\calk)$ defined in
\eqref{LocalCyclicReplicaTransformation}.
\end{itemize}
\end{proposition}

\begin{proof}[\bf Proof]
The relation \eqref{SpinTimesSpin} shows that $H_{\rm replica}$ has the form
\eqref{MultipleReplicaHamiltonian}. The invariance under the global replica
group follows by considering the effect on $H_{\rm replica}$ expressed in the
$\vec\sigma$ variables, where the transformation permutes the various terms
$H_\Lambda(\sigma^{(\alpha)})$ in the first expression for $H_{\rm replica}$ in
\eqref{MultipleReplicaHamiltonianSigma}.

In order to see that $H_{\rm replica}(\vec \sigma)$ is not invariant under the
local cyclic replica group, we give a configuration $\vec\sigma$ and set
$\calk$ that provides a counterexample in the case $n=2$.  It is easiest to
visualize this configuration by illustrating it; see the left side of Figure
\ref{Fig:NoLocalReplicaSymmetry}. We choose $\calk$ to be the centermost square
in the configuration (with $\sigma^{(1)}=+1$ and $\sigma^{(2)}=-1$), and choose
$\pi_\calk\in S^c_n(\calk)$ to flip the spins in $\calk$.  The action of
$\pi_\calk$ produces the configuration on the right side of the figure,
and it lowers the energy by $4\abs{\partial\calk}$.  In other words, $H_{\rm
replica}(\vec \sigma)-H_{\rm replica}(\pi_\calk\vec
\sigma)=4\abs{\partial\calk}$, showing that $H_{\rm replica}$ is not invariant
under the action of $S^c_n(\calk)$.

\begin{figure}[h]
\centering
  \psfrag{p}[cc]{$+$}
  \psfrag{m}[cc]{$-$}
\begin{tabular}{cccc}
    \raisebox{10mm}{$\sigma^{(1)}:$} & \includegraphics[scale=0.7]{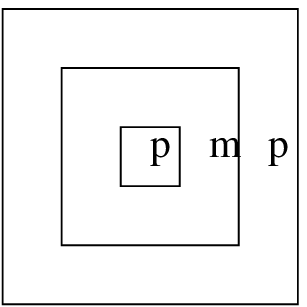} & \raisebox{10mm}{$\longrightarrow$} & \includegraphics[scale=0.7]{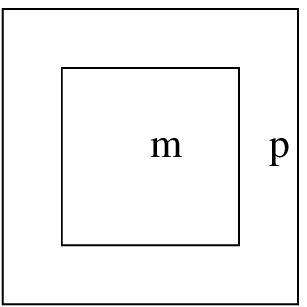} \\
    \raisebox{10mm}{$\sigma^{(2)}:$} & \includegraphics[scale=0.7]{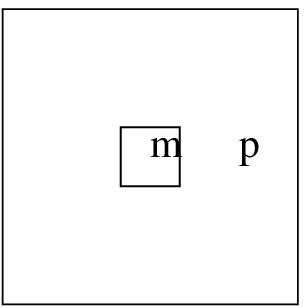} & \raisebox{10mm}{$\longrightarrow$} & \includegraphics[scale=0.7]{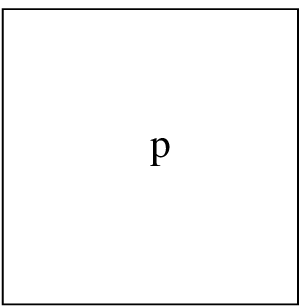}
\end{tabular}
\caption{A counter-example to local cyclic replica symmetry.
    \label{Fig:NoLocalReplicaSymmetry}}
\end{figure}

\end{proof}

\setcounter{equation}{0}
\section{Expectations} Define the expectation $\lraa{\ \cdot \ }\lb$ for the
replicated system as follows: for a function $F(\vec \sigma)$, let
    \be
        \lraa{F}\lb
        = \frac{1}{\fz} \sum_{\vec \sigma_i\atop i\in\Lambda} F(\vec \sigma)
            e^{-\beta H_{\rm replica}(\vec\sigma)}\;,
    \ee
where $\fz=\calz^n$, with $\calz$ is given in \eqref{NormalPartitionFunction}.
In case that $F(\vec \sigma)=f(\sigma^{(\alpha)})$ only depends on one
component $\sigma^{(\alpha)}$, the expectation $\lraa{\ \cdot\ }\lb$ reduces to
the expectation $\lra{\ \cdot\ }\lb$.  In this case
    \be
        \lraa{f(\sigma^{(\alpha)}) }\lb
        =  \lra{f(\sigma) }\lb\;,
        \qsp{for}
        \alpha=1,\ldots,n\;.
    \ee

We now introduce the generating function $S(\mu)$ for expectations of products
of spins. Let $\mu$ be a function from $\Lambda$ to $\C$ and let
    \be
        \sigma(\mu)
        = \sum_{i\in\Lambda} \mu_i\,\sigma_i\;,
        \qsp{and correspondingly}
        \sigma^{(\alpha)}(\mu)
        = \sum_{i\in\Lambda} \mu_i\,\sigma^{(\alpha)}_i\;.
    \ee
Then define
    \be
        S(\mu)
        = \lra{e^{\sigma(\mu)} }\lb
        = \lraa{e^{\sigma^{(\alpha)}(\mu)} }\lb\;.
    \ee
The expectations of $n$ spins are derivatives of the generating function,
    \be
        \lra{\sigma_{i_1}\,\sigma_{i_2}\cdots
        \sigma_{i_n}}\lb
        = \l. \frac{\partial^n}{\partial \mu_{i_1} \partial \mu_{i_2}
            \cdots\partial \mu_{i_n}}
        S(\mu) \r|_{\mu_i=0}
        = \lraa{\sigma^{(1)}_{i_1}\,\sigma^{(1)}_{i_2}\cdots
        \sigma^{(1)}_{i_n}}\lb\;.
        \label{Expectations}
    \ee
The expectations \eqref{Expectations} are $n$-multi-linear, symmetric,
functions of the spins,
    \be
        \lra{\sigma(\mu)^n)}\lb
        = \sum_{i_1,\ldots,i_n=1}^n \mu_{i_1}\cdots\mu_{i_n}
        \lra{\sigma_{i_1}\,\sigma_{i_2}\cdots
        \sigma_{i_n}}\lb\;.
    \label{Multilinearity}
    \ee
One can recover the expectation
$\lra{\sigma_{i_1}\,\sigma_{i_2}\cdots\sigma_{i_n}}\lb$ from the expectations
of powers of $\sigma(\mu)$ by polarization,
    \be
        \lra{\sigma_{i_1}\,\sigma_{i_2}\cdots
        \sigma_{i_n}}\lb
        = \frac{1}{2^{n}n!}\sum_{\ep_1,\ldots,\ep_n=\pm1} \ep_1\cdots\ep_n
        \lra{\lrp{\ep_1\sigma_{i_1}+\cdots+\ep_n\sigma_{i_n}}^n}\lb\;.
    \label{Polarization}
    \ee

\subsection{Truncated Expectations\label{Sect:TruncatedCorrelations}} The truncated
expectation of a product of $n$ spins is a generalization of the correlation of
two spins.  The truncated expectation vanishes asymptotically as one translates
any subset of the spin locations a large distance away from the others.

The generating function of the connected expectations is
    \be
        G(\mu)
        = \ln S(\mu)
        = \ln  \lra{e^{\sigma(\mu)}}\lb \;.
    \ee
One defines the truncated (connected) expectations as
    \beq
        \lra{\sigma_{i_1}\,\sigma_{i_2}\,\sigma_{i_3}\cdots
        \sigma_{i_n}}^{\rm T}\lb
%        &=& \l. \frac{\partial^n}{\partial \mu_{i_1} \partial \mu_{i_2}
%            \cdots\partial \mu_{i_n}}
%        \ln \lra{e^{\sigma(\mu)}}\lb \r|_{\mu_i=0}\nn
        &=& \l. \frac{\partial^n}{\partial \mu_{i_1} \partial \mu_{i_2}
            \cdots\partial \mu_{i_n}}
        G(\mu)\r|_{\mu_i=0}\;.
    \eeq
A standard representation of
$\lra{\sigma_{i_1}\,\sigma_{i_2}\,\sigma_{i_3}\cdots
        \sigma_{i_n}}^{\rm T}\lb$ in terms of sums of products of expectations
can be formulated in terms of the set $\calp$ of partitions of
$\{i_1,i_2,\ldots,i_n\}$. Suppose that a set $P\in\calp$ has cardinality
$\abs{P}$. Then
    \be
        \lra{\sigma_{i_1}\,\sigma_{i_2}\,\sigma_{i_3}\cdots
        \sigma_{i_n}}\lb
        = \sum_{\calp} \prod_{P\in\calp} \lra{\sigma^P}\lb^{\rm T}\;.
    \label{ExpectationsFromTruncated}
    \ee
Like the expectations \eqref{Expectations}, the $n$-truncated expectations
satisfy the $n$-multi-linear relation
\eqref{Multilinearity}--\eqref{Polarization}. Thus
    \be
        \lra{\sigma(\mu)^n}\lb^{\rm T}
        = \sum_{i_1,\ldots,i_n=1}^n \mu_{i_1}\cdots\mu_{i_n}
        \lra{\sigma_{i_1}\,\sigma_{i_2}\,\sigma_{i_3}\cdots
        \sigma_{i_n}}^{\rm T}\lb\;,
    \label{TruncatedMultilinearity}
     \ee
and
    \be
        \lra{\sigma_{i_1}\,\sigma_{i_2}\cdots
        \sigma_{i_n}}^{\rm T}\lb
        = \frac{1}{2^{n}n!}\sum_{\ep_1,\ldots,\ep_n=\pm1} \ep_1\cdots\ep_n
        \lra{\lrp{\ep_1\sigma_{i_1}+\cdots+\ep_n\sigma_{i_n}}^n}^{\rm T}\lb\;.
    \label{TruncatedPolarization}
    \ee

\subsection{Truncated Functions as Replica Expectations}
The form of the replica variables $\vec s$ leads to an elementary
representation of the truncated (connected) expectations of products of spins.
Ultimately we show that this yields exponential decay at low temperatures with
a rate governed by the length of the shorted tree-graph connecting all the
spins.  (A similar argument presumably works at high temperature.)

Our expansion method uses replica symmetry to arrange that each term in the
expansion either exhibits the desired decay rate, or else it is canceled by
other terms as a consequence of local cyclic replica symmetry. We begin by
establishing a known representation of the connected correlation of $n$ spins
as an expectation of $n$ replica variables  introduced above. This
representation was discovered by P. Cartier (unpublished); our presentation is
based on Sylvester's treatment \cite{Sylvester} using $s^{(1)}$.  Let $\gcd$
denote the greatest common divisor.

\begin{proposition} Let $\vec s$ be defined in \eqref{ReplicaSums}
with $n$ replica copies, and let $\gamma\in (1,\ldots,n)$ satisfy
$\gcd(n,\gamma)=1$. Then
    \be
        \eqbox{ \lra{\sigma_{i_1}\,\sigma_{i_2}\cdots
        \sigma_{i_n}}\lb^{\rm T}
        = n^{(n-2)/2}\,\lraa{s^{(\gamma)}_{i_1}\,s^{(\gamma)}_{i_2}\cdots
        s^{(\gamma)}_{i_n}}\lb}\;.
    \label{TruncatedCorrelationRepresentation}
    \ee
\end{proposition}

\begin{lemma} \label{Lemma:First}  For all $\gamma=1,\ldots,n$,
    \be
        \lraa{s^{(\gamma)}_{i_1}\,s^{(\gamma)}_{i_2}\cdots
        s^{(\gamma)}_{i_n}}\lb^{\rm T}
        = n^{-(n-2)/2}\,\lra{\sigma_{i_1}\,\sigma_{i_2}\cdots
        \sigma_{i_n}}\lb^{\rm T}\;.
    \label{TruncatedSRepresentation}
    \ee
\end{lemma}
\begin{proof}[\bf Proof]
Using the multi-linearity \eqref{TruncatedMultilinearity}, and its analog for
the expectations $\lra{\ \cdot\ }\lb$ and $\lraa{\ \cdot\ }\lb$ of the
truncated functions, we infer that
    \beq
        &&\hskip-.4in\lraa{s^{(\gamma)}_{i_1}\,s^{(\gamma)}_{i_2}\cdots
        s^{(\gamma)}_{i_n}}\lb^{\rm T}\nn
        && =
        n^{-n/2}\lraa{\sum_{\alpha_1,\ldots,\alpha_n=1}^n
        \omega^{\gamma\alpha_1+\cdots+\gamma\alpha_n-\gamma n}\,
        \sigma^{(\alpha_1)}_{i_1}\,\sigma^{(\alpha_2)}_{i_2}\cdots
            \sigma^{(\alpha_n)}_{i_n}}\lb^{\rm T}\nn
        &&= {n^{-n/2}} \sum_{\alpha_1,\ldots,\alpha_n=1}^n
        \omega^{\gamma\alpha_1+\cdots+\gamma\alpha_n-\gamma n}
        \;\lraa{\sigma^{(\alpha_1)}_{i_1}\,\sigma^{(\alpha_2)}_{i_2}\cdots
            \sigma^{(\alpha_n)}_{i_n}}\lb^{\rm T}\;.\nn
    \eeq
Since the different components of $\vec \sigma_i$ are independent, the
expectations on the right vanishes unless $\alpha_1=\cdots=\alpha_n$. In this
case the truncated expectation of each copy equals the truncated expectation of
the original spins, and the sum yields $n$ such terms.  Therefore
\eqref{TruncatedSRepresentation} holds as claimed.

\end{proof}

\begin{lemma}\label{Lemma:Second}  Let $k\gamma \neq0$ {\rm (modulo} $n${\rm )}.  Then
    \be
        \eqbox{
        \lraa{s^{(\gamma)}_{i_1}\,s^{(\gamma)}_{i_2}\cdots  s^{(\gamma)}_{i_k}}\lb
        =  0}\;.
    \ee
\end{lemma}

\begin{proof}
Expand the expectation
    \beq
        &&\hskip-.3in \lraa{s^{(\gamma)}_{i_1}\,s^{(\gamma)}_{i_2}\cdots
        s^{(\gamma)}_{i_k}}\lb\nn
        &&=
        \frac{1}{n^{n/2}} \sum_{\alpha_1,\ldots,\alpha_n=1}^n
        \omega^{\gamma\alpha_1+\cdots+\gamma\alpha_k-\gamma k}
        \lraa{\sigma^{(\alpha_1)}_{i_1}\,\sigma^{(\alpha_2)}_{i_2}\cdots
            \sigma^{(\alpha_n)}_{i_k}}\lb\nn
        &&= \frac{1}{n^{n/2}} \sum_{\alpha_1,\ldots,\alpha_n=1}^n
        \omega^{\gamma\alpha_1+\cdots+\gamma\alpha_k-\gamma k}
        \lraa{\sigma^{(\alpha_1-1)}_{i_1}\,\sigma^{(\alpha_2-1)}_{i_2}\cdots
            \sigma^{(\alpha_n-1)}_{i_k}}\lb\;.\nn
    \eeq
In the second equality, we use the symmetry of the expectation $\lraa{\ \cdot\
}\lb$ under the global cyclic replica group $S^c_n\ni\pi^0$. Therefore
    \be
        \lraa{s^{(\gamma)}_{i_1}\,s^{(\gamma)}_{i_2}\cdots
        s^{(\gamma)}_{i_k}}\lb
        = \omega^{\gamma k}\,\lraa{s^{(\gamma)}_{i_1}\,s^{(\gamma)}_{i_2}\cdots
        s^{(\gamma)}_{i_k}}\lb\;.
    \ee
As long as $\gamma k\neq0$ (modulo $n$), it is the case that $\omega^{\gamma
k}\neq1$. Therefore the expectation must vanish.
\end{proof}

\begin{proof}[\bf Proof of the Proposition] The relation
\eqref{ExpectationsFromTruncated} also holds for the replica expectations,
    \be
        \lraa{s^{(\gamma)}_{i_1}\,s^{(\gamma)}_{i_2} \cdots
        s^{(\gamma)}_{i_n}}\lb
        = \sum_{\calp} \prod_{P\in\calp} \lraa{s^{(\gamma)\,P}}\lb^{\rm T}\;.
    \label{ReplicaExpectationsFromTruncated}
    \ee
Because $\gcd(n,\gamma)=1$, it is the case that $k\gamma\neq0$ (modulo $n$) for
all $k=1,\ldots,n-1$.  Thus we can apply \Lref{Lemma:Second} to each such $k$,
and only the partition $P$ with all $n$ elements in one set survives in
\eqref{ReplicaExpectationsFromTruncated}. We infer
    \be
        \lraa{s^{(\gamma)}_{i_1}\,s^{(\gamma)}_{i_2}\,s^{(\gamma)}_{i_3}\cdots
        s^{(\gamma)}_{i_n}}\lb
        = \lraa{s^{(\gamma)}_{i_1}\,s^{(\gamma)}_{i_2}\,s^{(\gamma)}_{i_3}\cdots
        s^{(\gamma)}_{i_n}}\lb^{\rm T}\;.
    \ee
Using \Lref{Lemma:First} then completes the proof.

\end{proof}

\setcounter{equation}{0}
\section{Replica Condensation\label{Sect:ReplicaCondensation}}
In this section we investigate certain classes of configurations $\vec \sigma$
of the replica spins.  We see that for each class of configurations, there is a
local cyclic replica group (see \S \ref{Section:LocalReplicaGroup}) under which
the Hamiltonian $H_{\rm replica}$ of \eqref{MultipleReplicaHamiltonianSigma} is
invariant.  This leads to the phenomenon of {\em replica condensation} in which
all the spin localizations $i_1,\ldots,i_n$ must be localized within a given
region $\calk\subset\Lambda$ that we call a {\em continent.}

\subsection{Continents}
Each configuration of spins $\vec\sigma$ in the volume $\Lambda$  defines a
{\em sea} $\cals(\vec\sigma)$, surrounding a set of {\em continents}
$\calk(\vec\sigma)$.  The sea starts at the boundary boundary $\partial
\Lambda$ of the region $\Lambda$.  The boundary of a continent appears if any
one of the components of $\vec \sigma$ changes its value. Continents have a
substructure arising from the different configurations of the individual
components $\sigma^{(\alpha)}$ within the continent. We say more about this
substructure when defining {\em replica continent contours} in
\S\ref{Sect:ReplicaContinentContours}.  In the following we utilize the notion
of ``connectedness'' introduced in \S\ref{Sect:LatticeConnectedness}.

\begin{figure}[h]
    \centering
    \psfrag{B1}[cc]{$\calk_1$}
    \psfrag{B2}[cc]{$\calk_2$}
    \psfrag{B3}[cc]{$\calk_3$}
    \psfrag{B4}[cc]{$\calk_4$}
    \psfrag{B5}[cc]{$\calk_5$}
    \psfrag{S}[cc]{$\cals(\vec\sigma)$}

    \includegraphics{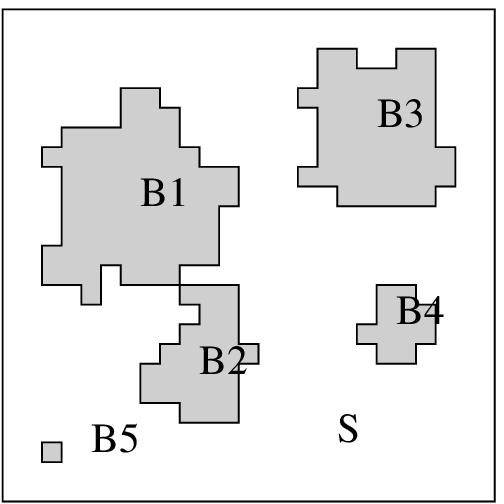}
    \caption{The set of continents
    $\calk(\vec\sigma)=\{\calk_1,\dots,\calk_5\}$ in the sea $\cals(\vec\sigma)$.
     \label{Fig:continents}}

\end{figure}

\begin{definition} \label{Defn:Sea_Continent_Island_Contour}
Consider a configuration $\vec \sigma$.  The {\em replica sea}
$\cals(\vec\sigma)$ is the connected component of the set $\{i\,|\
\vec\sigma_i=(+1,\ldots,+1)\}$ that meets the boundary $\partial \Lambda$ of
$\Lambda$.  The {\em continents} $\calk_j$  are the connected components of the
complementary set,
    $
        \cals^c(\vec\sigma)
        = \calk_1\cup\cdots\cup \calk_r\;.
    $
The set of continents $\calk(\vec\sigma)$ is
    \be
        \calk(\vec\sigma)=\{\calk_1,\ldots,\calk_r\}\;.
    \ee
\end{definition}
\noindent We illustrate this definition in Figure~\ref{Fig:continents}.

\subsection{Local Cyclic Replica Symmetry}
In \S\ref{Section:BasicReplicaSymmetry} we saw that a global replica symmetry
transformation leaves $H_{\rm replica}(\vec \sigma)$ invariant, and that a
local replica symmetry transformation does not necessarily do so.   We now
recover local cyclic replica symmetry by choosing the localization $\calk$ in
$S^c_n(\calk)$ to be a continent.

\begin{proposition} \label{Prop:LocalSymmetry}
Let $\calk\in \calk(\vec\sigma)$. Then the local cyclic replica group $
S^{c}_n(\calk)$ defined in \eqref{LocalCyclicReplicaTransformation} preserves
the continent $\calk$ and the Hamiltonian $H_{\rm replica}(\vec\sigma)$.  For
$\pi_\calk\in S^{c}_n(\calk)$,
    \be
        H_{\rm replica}(\vec\sigma)=H_{\rm replica}(\pi_\calk(\vec\sigma))\;.
    \label{CorrelationLemma1}
    \ee
\end{proposition}

\begin{proof}
The action of $ S^{c}_n(\calk)$ on $\vec\sigma$ leaves invariant spins
$\vec\sigma_i=(+1,\ldots,+1)$, so it changes neither the sea
$\cals(\vec\sigma)$ nor the definition of continents.  Hence it also does not
change the contribution of nearest neighbor spins to the energy either inside
or outside the continent.  The local permutation also does not alter the energy
across the island boundary, because all the components outside the island have
value $+1$ and are invariant under the permutation.
\end{proof}

\subsection{Symmetry Ensures Condensation}
We now establish the property of condensation. We use the representation
\eqref{TruncatedCorrelationRepresentation} for the truncated correlation
function of $n$ spins.  We may choose any $\gamma$ with $\gcd(n,\gamma)=1$, so
for simplicity we consider the case $\gamma=1$.

\begin{proposition}[\bf Condensation]  In the expectation
$\lraa{s^{(1)}_{i_1}\cdots s^{(1)}_{i_n}}\lb$, any configuration $\vec \sigma$
giving a nonzero contribution has all the sites $i_1,\ldots,i_n\in\calk$ lying in a single
continent $\calk\in\calk(\vec\sigma)$.
\label{Prop:Condensation}
\end{proposition}

\begin{lemma}
Consider a given configuration $\vec\sigma$ and a continent $\calk\in
\calk(\vec\sigma)$ containing at least one but not all the sites
$i_1,\ldots,i_n$.  Let $\pi_\calk^k$ denote
$\pi_\calk$ applied $k$ times. Then
    \beq
        &&\hskip-.7in\sum_{k=0}^{n-1}\lrp{\pi_\calk^k s_{i_1}^{(1)}}
            \cdots \lrp{\pi_\calk^k s_{i_n}^{(1)}}\;
            e^{-\beta H_{\rm replica}\left(\pi_\calk^k(\vec{\sigma})\right)}\nn
        && =
        \sum_{k=0}^{n-1}s_{i_1}^{(1)}\lrp{\pi_\calk^k(\vec{\sigma})}
            \cdots s_{i_n}^{(1)}\lrp{\pi_\calk^k(\vec{\sigma})}\;
            e^{-\beta H_{\rm replica}\left(\pi_\calk^k(\vec{\sigma})\right)}\nn
        &&=0 \;.
    \eeq
    \label{Lem:Condensation}
\end{lemma}

\begin{proof}
From \Pref{Prop:LocalSymmetry} we infer that the energy in the permuted
configuration is unchanged by the permutation,
    \be
        H_{\rm replica}\left(\pi_\calk^k(\vec{\sigma})\right)
        = H_{\rm replica} (\vec\sigma)\; .
    \ee
Therefore, we only need consider  the changes to the spins $s^{(1)}_{i_k}$. Let
$l=\abs{\{k|i_k\in \calk\}}$ denote the number of sites $i_1,\ldots,i_k$ that
lie in $\calk$; clearly $1\leq l<n$. According to
\Pref{Prop:CyclicReplicaSymmetryDiagonal}, the application of $\pi_\calk$ to
$s^{(1)}_{i}$ gives a phase $\omega$ for $i\in \calk$.  The sum equals
    \beq
         &&\hskip-.4in \sum_{k=0}^{n-1}\omega^{kl} s_{i_{1}}^{(1)}s_{i_{2}}^{(1)} \cdots s_{i_{n}}^{(1)}
                         e^{-\beta H_{\rm replica}(\vec{\sigma})} \nn
         &&= s_{i_{1}}^{(1)}s_{i_{2}}^{(1)} \cdots s_{i_{n}}^{(1)} e^{-\beta H_{\rm replica}(\vec{\sigma})}
         \sum_{k=0}^{n-1}\omega^{kl}
         = 0 \; .
    \eeq
\end{proof}

\begin{proof}[Proof of \Pref{Prop:Condensation}]
The expectation is
    \be \label{proof_condensation_expectation}
        \lraa{s^{(1)}_{i_1}\cdots s^{(1)}_{i_n}}\lb
        = \sum_{\vec\sigma} s^{(1)}_{i_1}\cdots s^{(1)}_{i_n}
        e^{-\beta H_{\rm replica}(\vec\sigma)}/ \fz \; .
    \ee
If $\vec\sigma$ is a configuration where some site $i_k$ lies in the sea $i_k
\in \cals(\vec\sigma)$ then the spin has the value of the boundary,
$s^{(1)}_{i_k}=0$. We also have $s^{(1)}_{i_k}= 0$, if $i_k\in \calk$ and all
the $\sigma^{(\alpha)}$ take the same values on $\calk$.

Therefore, the only contributing configurations have all the sites $i_k$ lying
in continents where $\pi_\calk$ actually yields new configurations. In this
case, the sum in \Lref{Lem:Condensation} is a sub-sum of
\eqref{proof_condensation_expectation}. According to the lemma the sum is only
nonzero if all or none of the $i_k$ lie in the contintent $\calk$.
\end{proof}

\begin{figure}
    \psfrag{p}[cc]{$+$}
    \psfrag{m}[cc]{$-$}
    \psfrag{o}[cc]{$0$}
    \psfrag{B1}[cc]{$\calk_1$}
    \psfrag{B2}[cc]{$\calk_2$}
    \psfrag{c1}[lc]{color 1}
    \psfrag{c2}[lc]{color 2}
    \psfrag{S}[cc]{$\cals(\vec\sigma)$}

    \centering

    \subfigure[The contours of $\sigma^{(1)}$]{\label{fig1a}
                    \includegraphics[scale=0.9]{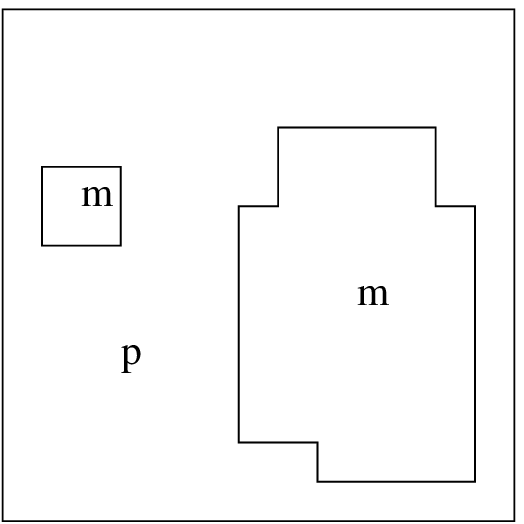}}
    \goodgap
    \subfigure[The contours of $\sigma^{(2)}$]{\label{fig1b}
                    \includegraphics[scale=0.9]{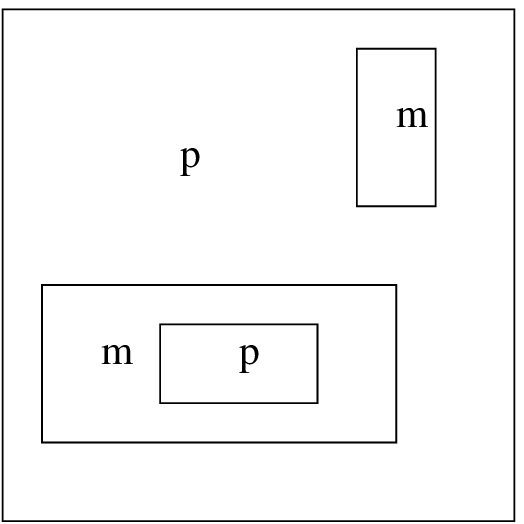}} \\

    \subfigure[The contours of $\vec\sigma$]  {\label{fig1c}
                    \includegraphics[scale=0.9]{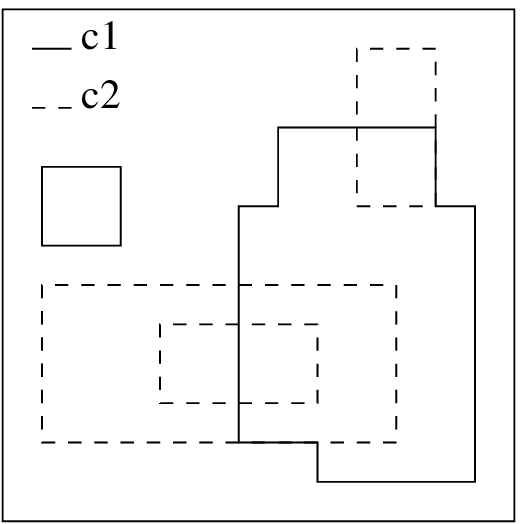}} \\
    \subfigure[The set of continents $\calk(\vec\sigma)=\{\calk_1,\calk_2\}$]
        {\label{fig1d}
                    \includegraphics[scale=0.9]{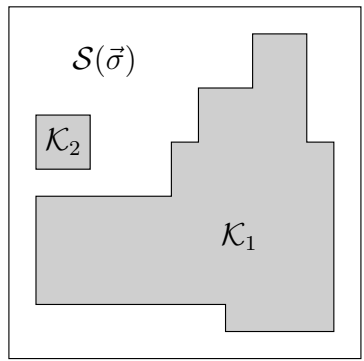}}
    \goodgap
    \subfigure[The continent contours $\vec C(\calk_1,\vec\sigma)$] {\label{fig1e}
                    \includegraphics[scale=0.9]{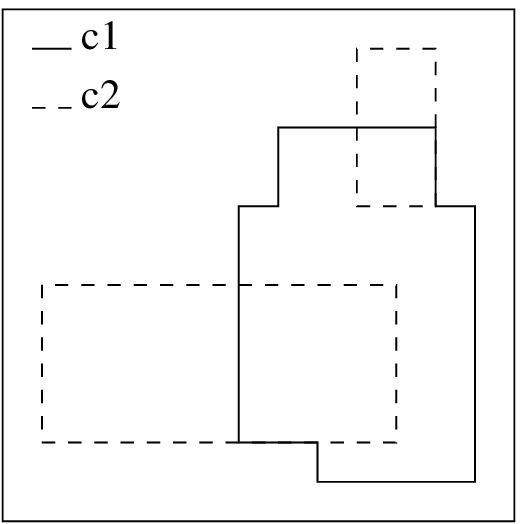}}

    \caption{An illustration of contours and continents in the case $n=2$.}
\end{figure}

\setcounter{equation}{0}
\section{\ Contours and the Energy\label{Sect:Contours}}
\subsection{\ Contours for Vector Spins $\vec\sigma$}
For each component $\sigma^{(\alpha)}$ of the vector spin, we can define
contours in the usual statistical mechanics sense.  These contours are the
boundaries between islands with different values of $\sigma^{(\alpha)}$,
as defined in \S \ref{Sect:LatticeConnectedness}. They are subsets of the lattice dual to $\Z^d$, consisting of
$(d-1)$-faces of $d$-cubes.

The $\vec\sigma$ contours are the direct sum of contours in the individual
components. In order to picture the boundaries of $\vec\sigma$, we assign
colors to the different components, corresponding to the label $\alpha$ used
above.  We illustrate these contours for a particular configuration in the case
$n=2$ in Figure~\ref{fig1a}--Figure~\ref{fig1c}.

\subsection{\ Replica Continent Contours\label{Sect:ReplicaContinentContours}}
Here we define appropriate {\em replica continent contours}
$\vec{C}(\calk,\vec{\sigma})$ in order to analyze the probability $\Pr(r)$ of
the occurrence of configurations containing a continent with a contour of
length $r$. We do not define $\vec C$ as the boundary $\partial \calk$.  The
problem is: while this boundary is a contour for $\vec\sigma$, it is not
necessarily  a contour for a component $\sigma^{(\alpha)}$. Usually $\partial
\calk$ consists of segments of contours of the components.

To estimate $\Pr(r)$ we use the relation between the
configuration $\vec\sigma$ with the replica continent contour $\vec{C}(\calk,\vec\sigma)$ and a configuration
$\vec\sigma^*$ with the contour removed.  This transformation removes all the contours of the component spins
$\sigma^{(\alpha)}$ that contribute to $\partial \calk$.  With this motivation,
we now give the appropriate construction.

\begin{Def} \label{Defn:ContinentContour}
For $\calk\in \calk(\vec\sigma)$ define the {\em replica continent contour} of
$\calk$ in the configuration $\vec\sigma$ as the vector $\vec{C}(\calk,\vec{\sigma})$
with components
    \be
         C^{(\alpha)}(\calk,\vec{\sigma})
        = \textrm{union of contours } C \textrm{ for }
        \sigma^{(\alpha)}\textrm{ with } \abs{C\cap\partial
        \calk}\neq 0
        \;,
    \ee
where $\abs{\ \cdot\ }$ is the measure of $(d-1)$-surfaces. This is the subset
of contours for $\vec\sigma$ meeting the boundary of the continent $\partial
\calk$.
\end{Def}

See the example in Figure~\ref{fig1e}. In a generic configuration, these
contours touch the boundary and penetrate arbitrarily into the
interior of the continent.

Several different configurations of the spin $\vec\sigma$ may have different
contours, but a common continent $\calk$.   Define the set of
possible contours for the continent $\calk$ as
    \beq
         C(\calk)
        & = & \left\{ \vec C(\calk,\vec{\sigma})\,|\
        \hensp{where}%\vec\sigma\textrm{ such that }
        \calk\in \calk(\vec{\sigma})\right\} .
    \label{SetContinentContours}
    \eeq
Finally, the length of any contour $\vec C\in  C(\calk)$ is just the sum over
the length of the constituent contours,
    \beq
        \abs{\,\vec C}
        =\sum_{\alpha=1}^{n} \abs{\,C^{(\alpha)}}\;.
    \eeq
With these definitions it is obvious that removing $\vec C(\calk,\vec\sigma)$
in the configuration $\vec\sigma$ is well-defined.  We just remove the
respective contours $C^{(\alpha)}(\calk,\vec\sigma)$ for the components
$\sigma^{(\alpha)}$, by flipping the sign of all the spins inside these
contours.

\begin{Def} \label{Defn:Flips}
For a configuration $\vec\sigma$ and a continent $\calk\in \calk(\vec\sigma)$,
write $\vec\sigma^{\,*}$ for the configuration where the contour $\vec
C(\calk,\vec\sigma)$ for the continent
 has been removed as described above.
\end{Def}
As a consequence of the removal of the replica continent contour the energy
$H_{\rm replica}$ is decreased by two times the length of the removed contours.
This the generalization of the fact that for each component spin, the energy is
given by two times the total length of the contours,
    \be
        H_{\rm replica}(\vec\sigma^{\,*})
        =H_{\rm replica}(\vec\sigma)-2\abs{\vec C(\calk,\vec\sigma)}\;.
    \label{SpinFlipHamiltonian}
    \ee

\setcounter{equation}{0}
\section{\ \ Counting Random Surfaces in $\R^d$}
In order to prove the tree decay we need an exponential bound on the number of
possible connected contours.  These are surfaces in $\R^d$ composed of $r$
faces, each a unit $(d-1)$-cube. We call these {\em random surfaces} and prove
a bound that holds for general connected unions of faces, as defined in \S
\ref{Sect:LatticeConnectedness}. We also use the term {\em adjacent faces} as
in that section, to indicate that two faces share a $(d-2)$-dimensional
cube.

\begin{definition}
Let  $N(r)$ denote the number of connected, random surfaces of dimension
$(d-1)$, which contain exactly $r$ faces, including a given face
$S_0$.
\end{definition}

\begin{Prop} \label{Prop:EntropyBound}
There is a constant $a$ (independent of dimension) such that for $k_d=a2^d$,
    \be
        N(r) \leq k_d^r\;.
    \label{SurfaceCountingBound}
    \ee
\end{Prop}

\begin{figure}[h]
    \psfrag{p1}[cc]{$p_1$}
    \psfrag{p2}[cc]{$p_2$}
    \psfrag{p3}[cc]{$p_3$}
    \psfrag{p4}[cc]{$p_4$}
    \psfrag{p5}[cc]{$p_5$}
    \psfrag{p6}[cc]{$p_6$}
    \psfrag{S0}[cc]{$S_0$}

    \centering

    \includegraphics[scale=.7]{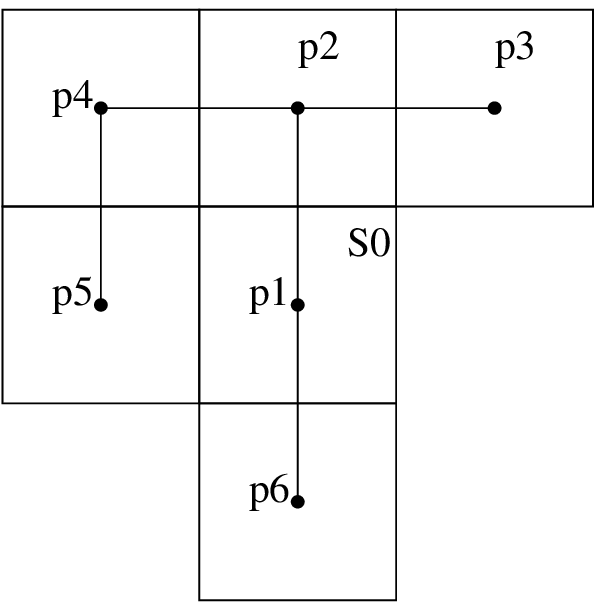}
    \qquad \qquad
    \includegraphics[scale=.7]{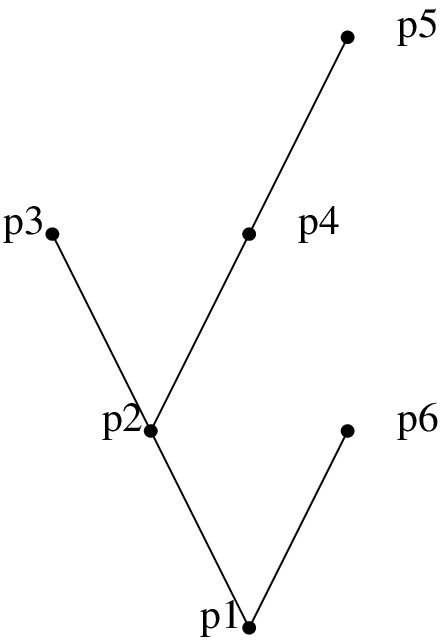}

\caption{A surface covered by a tree graph and the corresponding tree graph
rooted at the center of $S_0$. To simplify the illustration, all the angles
between the faces are set to $180^\circ$, while in general they may equal
$90^\circ$, $180^\circ$, or $270^\circ$. \label{Fig:CoveringTree}}

\end{figure}

\begin{proof}
The idea of the proof is to map each connected surface onto a rooted
tree-graph, whose edges connect the centers of adjacent faces of the surface,
and which touches each face. We say that the graph {\em covers} the surface.
One then counts the number of possible surfaces that can correspond to one
graph. The product of the number of possible tree graphs, times the number of
surfaces per graph, gives our bound.

The tree graph will have length $r$ and $r-1$ edges; the root of the tree is
the center $p_1$ of $S_0$, see Figure \ref{Fig:CoveringTree}.  The first branch
of the tree connects the root $p_1$ to the center $p_2$ of a face adjacent to
$S_0$. From there we draw another edge connecting to the center of a new
adjacent face (but we do not return to $p_1$).  If all the adjacent surface
elements are already in the tree graph, we cannot continue this branch. At this
point we move in the reverse direction along the branch, face by face, until we
reach a face having an adjacent face that is not yet covered by the
tree.  Starting at this place we start a new branch. We continue in this manner
until we cover the entire surface.

In this manner we assign at least one tree diagram to every connected surface.
This also means that every possible connected surface with $r$ faces can be
constructed by choosing a tree graph and attaching new faces in the order given
by the tree structure. The number of planar tree graphs with $r-1$ edges is the
Catalan number $C_{r-1}$, see example 6.19.e of Stanley \cite{Stanley2}. Hence
    \be
        C_{r-1}=\frac{1}{r} \binom{2(r-1)}{r-1} \leq (2e)^r\;,
    \ee
where the bound follows from the elementary inequality
    \be
        \binom{v}{w}
        \leq \left(\frac{ev}{w}\right)^w\;.
    \label{Elem1}
    \ee

An upper bound for the number of ways to add a single face as one builds up the
surface along the tree-graph is $2^{d-1}\cdot 3$. A face has $2^{d-1}$ sides to
attach an adjacent face, and every attachment can be done with one of the
angles $90^\circ$, $180^\circ$, or $270^\circ$. Therefore we infer the bound
\eqref{SurfaceCountingBound} with $a=3e$, namely
    \beq
        N(r)
        & \leq & (2^{d-1} \cdot 3)^r  C_{r-1} \nn
        & \leq & (2^{d-1} \cdot 3 \cdot 2e)^r
        = k_d^r.
    \eeq
\end{proof}
\goodbreak

\setcounter{equation}{0}
\section{\ \ \ Tree Decay\label{Sect:TreeDecay}}
%%%%%%%%%%%%%%%% EDIT HERE
In this section we prove the decay bound for the truncated correlation
functions.  We base the proof on condensation.  Starting from the
representation \eqref{TruncatedCorrelationRepresentation}, namely
    \be
        \lraa{ s_{i_{1}}^{(1)}\cdots s_{i_{n}}^{(1)} }\lb
        = n^{-(n-2)/2}\langle \sigma_{i_{1}}\cdots
                        \sigma_{i_{n}}\rangle^{\mathrm{T}}\lb\;,
%             & = & \sum_{\vec{s}}s_{i_{1}}^{(1)}s_{i_{2}}^{(1)}\cdots s_{i_{n}}^{(1)}e^{-\beta H(\vec{s})}
    \label{TruncatedFunctionRepn}
    \ee
we use the fact established in \Pref{Prop:Condensation} that every
non-vanishing contribution contains a continent $\calk$ with all the points
$i_1,\ldots,i_n$.
\begin{prop}  There are constants $a,b$ depending on $d$, but independent of $\Lambda$,
such that if $1\le \delta = \beta - b \ln n$ (hence requiring $\beta\ge
\beta_n= O(\ln n)$), then the truncated correlation functions satisfy
    \be
        \abs{\langle \sigma_{i_{1}}\cdots
                        \sigma_{i_{n}}\rangle^{\mathrm{T}}\lb}
        \leq a \,n^n\,e^{-\delta\tau(i_{1},\dots,i_{n})}\;.
    \label{TreeDecayBound}
    \ee
Here $\tau(i_{1}, \dots, i_{n})$ is the length of the shortest tree connecting
$i_{1}$, $\dots$, $i_{n}$.
    \label{Prop:CorrelationDecay}
\end{prop}

\subsection{\ \ Outline of the Proof}
We have shown in \Pref{Prop:Condensation} that each non-vanishing contribution
to the expectation \eqref{TruncatedFunctionRepn} contains a
condensate continent $\calk$ containing all the points $i_1,\ldots,i_n$.  As a
consequence, every possible replica contour $\vec C\in C(\calk)$ has minimal
length $\tau(i_{1},\dots,i_{n})$.

We formulate the sum over configurations
    \be
       \lra{\sigma_{i_{1}}\cdots
                        \sigma_{i_{n}}}^{\mathrm{T}}\lb
        = n^{(n-2)/2} \frac1\fz \sum_{\vec\sigma} s_{i_{1}}\cdots
                        s_{i_{n}} e^{-\beta H_{\rm replica}(\vec\sigma)}\;,
    \label{TruncatedCorrelationRepn}
    \ee
as a sum over configurations with contours $\vec C$ of length $r$ and a sum
over $r$. We claim that the probability $\Pr(r)$ that a replica contour $\vec
C$ occurs with $\abs{\vec C}=r$ satisfies the bound
    \be
        \Pr(r)
        \le e^{-\beta \abs{\vec C}}
        = e^{-\beta r}\;.
    \ee
To complete the proof we use the entropy bound \Pref{Prop:EntropyBound}, along
with an estimate on the number of configurations that contain a given contour
$\vec C$. These estimates, together with the fact that $\abs{s^{(1)}_i}\le
n^{1/2}$,  yield the desired bound.  We now break the proof into a sequence of
elementary steps.

\subsection{\ \ Details of the Proof}
\paragraph{\bf Rewrite the Sum:}
Consider the sum \eqref{TruncatedCorrelationRepn}, with the restriction of
\Pref{Prop:Condensation}. Recall that the replica continent borders $\vec
C=\vec C(\calk,\vec\sigma)$, and the set of configurations containing such a
replica continent $C(\calk)\ni\vec C(\calk,\vec\sigma)$ is given in
\Dref{Defn:ContinentContour}. One can rewrite the sum as an iterated sum,
    \be
        \sum_{\vec \sigma}
        = \sum_{r=\tau(i_1,\dots,i_n)}^\infty \;\,
                \sum'_{\calk,\vec C}\;\,
                \sum''_{\vec \sigma}\;.
    \ee
For fixed $\calk$ and $\vec C$, the sum $\sum''$ denotes the sum over
configurations containing the continent $\calk\in\calk(\vec \sigma)$ with the
continent border $\vec C=\vec C(\calk,\vec \sigma)$,
    \be
        \sum''_{\vec \sigma}
        = \sum_{ \scriptsize \begin{array}{c}
        \vec{\sigma}
        \textrm{ with }\calk\in \calk(\vec{\sigma})\;,\;
            \vec C=\vec C(\calk,\vec{\sigma}) \end{array} }\;.
    \label{SumPrime2}
    \ee
The sum $\sum'$ ranges over the possible continents $\calk$ containing the $n$
sites $i_1,\ldots,i_n$, and their possible borders $\vec C$ of length
$\abs{\vec C}=r$.  Thus
    \be
        \sum'_{\calk,\vec C}
        = \sum_{ \calk \supset \left\{ i_{1},\dots,i_{n}\right\} }
             \sum_{ \scriptsize \begin{array}{c}
             \vec C\in C(\calk) \\ \textrm{with } \left|\vec C\right|=r \end{array} }\;.
    \label{SumPrime1}
    \ee
Finally we sum over $r$, which is bounded from below by the minimal size
$\tau(i_1,\ldots,i_n)$.

One interprets the sum $\sum''$ as the {\em  energy} contribution to the sum,
namely the probability
    \be
        \Pr(r)
        = \frac1\fz \sum'' e^{-\beta H(\vec\sigma)} \; ,
    \ee
for the states $\vec{\sigma}$ with $\calk\in \calk(\vec{\sigma})$.  Likewise
one interprets the sum $\sum'$ as the {\em entropy} contribution to the sum.
Define the {\em entropy factor} $\caln(r)$ by
    \be
        \caln(r)=\sum'_{\calk, \vec C} 1\;.
    \ee
The entropy counts the number of different shapes for $\vec C$.

Using $\abs{\sigma_i}=1$, one has $\abs{s^{(1)}_i}\le n^{1/2}$.  Thus we obtain
the bound
    \beq
        \abs{\langle \sigma_{i_{1}}\cdots
                        \sigma_{i_{n}}\rangle^{\mathrm{T}}\lb}
        & =    & n^{(n-2)/2} \l| \lraa{s_{i_1}^{(1)} \cdots s_{i_n}^{(1)} }\lb \r| \nn
        & \leq & n^{(n-2)/2} \frac1\fz\sum_{r=\tau}^\infty \sum'_{\calk,\vec C} \sum''_{\vec \sigma}
                 \l| s_{i_1}^{(1)} \cdots s_{i_n}^{(1)} \r|
                 e^{-\beta H(\vec{\sigma})} \nn
        &\le& n^{(n-2)/2} \sum_{r=\tau}^{\infty} n^{n/2} \caln(r)\, \Pr(r)\nn
        &=& n^{n-1}\sum_{r = \tau}^\infty \caln(r)\Pr(r)\;.
    \label{FinalOverallBound}
    \eeq
In the following we prove bounds on $\Pr(r)$ and on $\caln(r)$ that depend only
on $r$, on $\beta$, and on the dimension $d$.

\paragraph{\bf Bound the Entropy:}
We show that there are constants $A,B$ depending only on $d$ such that
$\caln(r)$ satisfies the exponential bound,
    \be
        \caln(r)
        \le A B^r n^r\;.
    \label{Entropy}
    \ee

We obtain this result by constructing the border contour $\partial \calk$ and
attaching $l$ colored sub-contours.  In this way one constructs any possible
$\vec C$ satisfying the conditions above.  The geometry of the contour (which
must surround $i_1$) requires that the starting face we choose in constructing
$\vec C$ must lie in a cube of side-length $(r-1)$, centered at $i_1$. Such a
cube contains at most $d r^d$ possible starting faces.

Using \Pref{Prop:EntropyBound}, the number of possible border contours is less
than $d r^d N(r)\le d r^d k_d^r$.  We now build up the full contour $\vec C$ by
attaching at least one and at most $r$ subcontours to $\partial \calk$ to
obtain the total number of faces $r$.  This can be done in a number of ways.
For $l$ sub-contours, the number of ways is  bounded by the product of
combinatorial factors:
    \[
    \begin{array}{cl}
        r^l  & \textrm{for the starting faces on $\partial \calk$,}\\
        k_d^r  & \textrm{for the shapes,}\\
        n^l  & \textrm{for the colors,}\\
        \binom{r-1}{l-1} & \textrm{for the lengths,}\\
        1/l! & \textrm{as the ordering of the subcontours is irrelevant.}
    \end{array}
    \]
Therefore
    \be
        \caln(r)
        \leq d r^d k_d^r
            \sum_{l=1}^r r^l k_d^r n^l \binom{r-1}{l-1} \frac{1}{l!}\;.
    \ee
We use the elementary inequalities
    \be
        r^d\le d!e^r\;,\qsp{and}
        \binom{r-1}{l-1} \leq 2^r\;.
%        \leq \left(\frac{ev}{w} \right)^w\; , \qsp{and}
%        \left(\frac{v}{w} \right)^w
%        \le \binom{v}{w}\;.
    \label{Elem2}
    \ee
Then
    \be
        \caln(r)
        \leq
            dr^d k_d^{2r} n^r \sum_{l=1}^r \frac{r^l}{l!} \binom{r-1}{l-1}
        \leq
           d d! \lrp{2e^{2} k_d^{2}}^r n^r\;.
    \ee
This bound has the form \eqref{Entropy} with $A=dd!$ and $B=2e^{2} k_d^{2}$.

\paragraph{\bf Bound the Energy Factor:}
The energy bound has the form
    \be
        \Pr(r) \leq e^{-\beta r},
    \ee
where $\calk$ (implicitely contained in $\sum'$) is any fixed connected set
with $\left\{ i_{1}, \dots, i_{n} \right\} \subset \calk$ and $\vec C \in
C(\calk)$ is any fixed extended border with $\left| \vec C \right| = r$. The
idea is to compare every summand in the numerator to a summand in the
denominator. For any given $\vec\sigma$ with $\calk\in \calk(\vec\sigma)$ and
$\vec C(\calk,\vec\sigma)=\vec C$, we can take away the contours in $\vec C$
obtaining the unique $\vec\sigma^{*}$ as described in
\Dref{Defn:Flips}. Because of the difference in energy this gives an additional
factor $e^{-\beta r}$ for the term in the numerator. As the procedure works for
all the summands, we infer
    \be
    \Pr(r) =    \frac{\sum''_{\vec \sigma} e^{-\beta H(\vec{\sigma})}}
                      {\sum_{\vec{\sigma}} e^{-\beta H(\vec{\sigma})}}
            \leq \frac{\sum''_{\vec \sigma} e^{-\beta H(\vec{\sigma}^{*})} e^{-\beta r}}
                      {\sum''_{\vec \sigma} e^{-\beta H(\vec{\sigma}^{*})}}
            =    e^{-\beta r}.
\ee

\paragraph{Tree Decay:}
The bound \eqref{TreeDecayBound} now follows.  Using \eqref{FinalOverallBound},
one has
    \beq
    \l| \langle \sigma_{i_1}\cdots\sigma_{i_n} \rangle ^{\mathrm{T}}\lb \r|
        & \leq & n^{n-1}\sum_{r = \tau}^\infty \caln(r)\Pr(r)
        \le
            n^{n-1} \sum_{r=\tau}^\infty AB^r n^r e^{-\beta r} \nn
        &=& n^{n-1} A\sum_{r=\tau}^\infty  e^{-(\beta-b\ln n) r}\;,
    \eeq
where $b=\ln B$ and where $\tau=\tau(i_{1},\dots,i_{n})$. The last sum
converges for $\beta>b \ln n$.  With $1\le\delta_n=\beta-b\ln n$, this gives
    \be
         \abs{\langle \sigma_{i_1}\cdots\sigma_{i_n} \rangle ^{\mathrm{T}}\lb}
         \le  n^{n-1} A (1-e^{-\delta_n})^{-1} \;e^{-\delta_n \tau}\;.
    \ee
As $1\le\delta_n $, one can take
    \be
        a= A \,(1-e^{-\delta_n})^{-1}
        \le Ae(e-1)^{-1} \;.
    \ee
This completes the proof of \Pref{Prop:CorrelationDecay}. $\hfill\square$

%% ... AND HERE ++++++++++++++++++++++++++++++++++++++++++++++++++++++++++++++++++++++++++++++++++++++++++++++++++++++++++

\bibliographystyle{plain}
\bibliography{Tree_Decay}

\def\cprime{$'$}
\begin{thebibliography}{1}

\bibitem{Stanley2}
Richard Stanley.
\newblock {\em Enumerative Combinatorics, Volume 2}.
\newblock Cambridge University Press.
\newblock Cambridge Studies in Advanced Mathematics 62, Cambridge, New York
  1999.

\bibitem{Sylvester}
Garrett Sylvester.
\newblock Representations and inequalities for {I}sing model {U}rsell
  functions.
\newblock {\em Comm. Math. Phys.}, 42:209--220, 1975.

\end{thebibliography}
\goodbreak
\end{document}